# Electromechanical properties of freestanding graphene functionalized with tin oxide (SnO$_2$) nanoparticles

L. Dong,[1,2,a] J. Hansen,[2] P. Xu,[3] M. L. Ackerman,[3] S. D. Barber,[3] J. K. Schoelz,[3] D. Jun,[3] and P. M. Thibado[3,b]

[1]*College of Materials Science and Engineering, Qingdao University of Science and Technology, Qingdao 266042, China*

[2]*Department of Physics, Astronomy, and Materials Science, Missouri State University, Springfield, Missouri 65897, USA*

[3]*Department of Physics, University of Arkansas, Fayetteville, Arkansas 72701, USA*

Freestanding graphene membranes were functionalized with SnO$_2$ nanoparticles. A detailed procedure providing uniform coverage and chemical synthesis is presented. Elemental composition was determined using scanning electron microscopy combined with energy dispersive X-ray analysis. A technique called electrostatic-manipulation scanning tunneling microscopy was used to probe the electromechanical properties of functionalized freestanding graphene samples. We found ten times larger movement perpendicular to the plane compared to pristine freestanding graphene, and propose a nanoparticle encapsulation model.

---

[a] Electronic mail: lifengdong@MissouriState.edu
[b] Electronic mail: thibado@uark.edu.

Solar cells utilizing solid-state semiconductor materials such as Si have been studied for over 50 years and are approaching the theoretical power conversion efficiency limit of 30%. Nevertheless, they are comparatively still too expensive for mass production. In the past ten years polymer heterojunction solar cells, which use cheaply manufactured organic polymers as electron donors placed in contact with an electron acceptor, have emerged as one of the leading candidates for the next generation of solar cells[1-3]. First made in 1992, polymer/fullerene ($C_{60}$) blends represented a large step forward with efficiencies of up to 2.5%[4-6]. Due to their unique electronic structure, fullerenes are excellent electron acceptors, and they are easily dispersed into a donor medium, leading to improved charge separation and hindered charge recombination compared to other polymer/donor junctions[7,8]. Single-walled carbon nanotubes promise even more efficient conversion due to their potentially large surface area and superior conductivity. In fact, polymer/nanotube junctions have been manufactured with efficiencies approaching 5%[8].

Another area of improvement can come from functionalizing the electron acceptor in the heterojunction cell to increase its electron affinity. This enhances charge separation and thus increases efficiency. The functional groups can be organic and covalently bonded to the donor and acceptor[7], or n-type inorganic nanocrystals formed from materials such as ZnO, $TiO_2$, $SnO_2$, etc[2,3]. In particular, it has been proven possible to uniformly functionalize carbon nanotubes with nanocrystals of $SnO_2$ by a chemical solution route[9]. Nanotubes have proved difficult to work with, however, as their electronic properties can vary widely depending on their morphology and because they tend to clump together due to strong intermolecular van der Waals interactions. Graphene, a single layer of carbon atoms and the 2D analog to fullerenes and nanotubes, promises to provide the advantages and avoid the disadvantages of carbon nanotubes[10]. Recently, the uniform deposition of $SnO_2$ nanocrystals onto graphene by simultaneous reduction of graphene oxide and oxidation of $Sn^{4+}$ by dimethyl sulfoxide was reported[11].

In this letter, we report about freestanding graphene and its functionalization with $SnO_2$ nanoparticles. Our approach is a relatively simple, two-step solution-based processing technique originally developed for carbon nanotubes, in which the nanoparticles are directly deposited on inexpensive, commercially available, freestanding graphene. Films are characterized using X-ray energy dispersive spectrometry (EDS) and field emission scanning electron microscopy (FESEM) to determine elemental composition and density of coverage. In addition, a specialized technique called electrostatic-manipulation scanning tunneling microscopy (EM-STM) is employed to probe the mechanical and electrostatic properties of the functionalized graphene compared to its pristine counterpart.

Graphene layers grown using chemical vapor deposition (CVD)[12,13] were transferred from Ni onto a 2000-mesh, ultrafine Cu grid with a square lattice of holes with sides measuring 7.5 μm in between support bars measuring 5 μm in width. $SnO_2$ nanoparticles were synthesized on the graphene surface via a chemical-solution route as illustrated in Fig. 1.[9] In brief, 100 mg of tin (II) chloride ($SnCl_2$) was dispersed in 10 mL of deionized water, and 175 μL of hydrochloric acid (HCl, 38%) was added to the mixture. The solution was then sonicated for 10 minutes. After sonication, the Cu grid with graphene layers was submerged in the solution for 60 minutes. Finally, the grid was taken out, washed with deionized water, and dried in an oven at 70 °C for 12 hours.

The morphology of $SnO_2$ nanoparticles formed on the graphene layers was examined using an FEI Quanta 200 FESEM equipped with a scanning transmission electron microscopy (STEM) detector and an Oxford INCA 250 silicon drift X-ray EDS. Since Sn has a much higher atomic number ($Z = 50$), high angle annular dark field (HAADF) STEM, a Z-contrast technique, was employed to investigate the distribution of SnO$_2$ nanoparticles on the graphene surface.[14,15]

The SEM and EDS findings are displayed in Fig. 2. First, a low magnification SEM image of pristine freestanding graphene on a Cu grid is shown in Fig. 2(a). Most of the holes in the mesh are fully covered by graphene, and we estimate 90% coverage. A close-up view is given in Fig. 2(b) to more clearly show the structure of the suspended graphene in a region of partial coverage. Next, a large scale image of freestanding graphene after SnO$_2$ functionalization is presented in Fig. 2(c). As before, yet surprisingly, the surface remains almost entirely covered with graphene due to its strength. The SnO$_2$ nanoparticles are distributed on the graphene surface in a uniform manner, as shown in Fig. 2(d), and the size of the nanoparticles ranges from 1 nm to 6 nm[9]. The EDS spectrum is shown in Fig. 2(e). The peaks for Sn ($L\alpha_1$, $L\beta_1$, and $L\gamma_1$ from ~3.5 keV to 4.0 keV) and O (~0.5 keV) further confirm the constituents of the nanoparticles, and their relative intensities are consistent with the expected composition. The signature of C (~0.3 keV) is from graphene, while the Cu (~0.9 keV) and Al (~1.5 keV) signals result from the Cu grid and Al sample holder, respectively.

EM-STM measurements were obtained using an Omicron ultrahigh-vacuum (base pressure is $10^{-10}$ mbar), low-temperature STM operated at room temperature. The samples were mounted on a flat tantalum sample plate using silver paint and loaded into the STM chamber via a load lock. The STM tips were electrochemically etched from 0.25 mm diameter tungsten wire via a custom double lamella setup with an automatic cutoff.[16] After etching, the tips were gently rinsed with distilled water, briefly dipped in a concentrated hydrofluoric acid solution to remove surface oxides,[17] and then loaded into the STM chamber. EM-STM measurements were taken with the feedback electronics left on, meaning that the tunneling current was maintained at a constant setpoint. The bias voltage between the tip and the grounded sample was then varied while the height change of the STM tip was recorded (this is similar to constant-current scanning tunneling spectroscopy). For comparison purposes EM-STM measurements were also taken on Au and on pristine freestanding graphene (i.e., before functionalization).

Since EM-STM is not a commonly used technique, measurements were first completed on an Au substrate, and the results are shown in Fig. 3(a). The three different height profiles correspond to feedback setpoint currents of 0.01 nA, 0.10 nA, and 1.00 nA. As the tip bias is increased from 0.1 V to 3.0 V, the height of the tip increases 1.5-2.0 nm depending on current. This is a feedback-on spectroscopy measurement and can be used to obtain an estimate for the local work function of the Au sample.[18] The data exhibit similar behavior for each setpoint current, but on average the height decreases with increasing current. This is consistent with the reduced tunneling barrier height required for a higher current in STM.

Even though the EM-STM procedure for the Au measurement is the same as for freestanding graphene, the results are fundamentally different. To illustrate, a schematic for an EM-STM measurement acting on the freestanding graphene is shown in the top region of Fig. 3(b). This schematic highlights the concept that as the applied bias increases, the freestanding graphene deforms toward the STM. The deformation is a result of the electrostatic attraction between the biased tip and grounded sample. The electrostatic force increases with increasing voltage.[19] Therefore, since the STM feedback circuit is left on, the tip will retract in order to maintain a constant tunneling current. This movement will continue until an elastic restoring force builds up in the graphene. Overall, when a sample is free to move, the recorded motions are typically much larger than stationary samples and contain information about the electromechanical properties of the material.

The EM-STM data obtained for pristine freestanding graphene is presented in Fig. 3(c). Similar to the Au sample, these measurements were acquired for setpoint currents of 0.01 nA, 0.10 nA and 1.00 nA. Over the same voltage range, the height change of the STM tip is now 5-25 nm, or a factor of ten greater than the Au. Note, the actual tunneling current is simultaneously recorded (not shown), and it remained constant (within a few percent) throughout the measurement. The current can be constant only if the graphene is following the tip as it retracts; otherwise the current would exponentially decrease with increasing height. It is interesting to also notice that the height dramatically increases with increasing current (opposite of the Au). This is because the electrostatic force, in addition to being directly dependent on bias voltage, is also influenced by the tunneling current. Basically, in order to achieve larger currents, the tip must move slightly closer to the sample, thereby greatly increasing the electrostatic attraction between them.

In order to better quantify the electrostatic interaction between the tip and sample, we developed the following model. The STM tip is treated as a biased conducting sphere with radius 20 nm, and the graphene is treated as an infinite grounded conducting plane held initially 0.5 nm from the surface of the sphere. The force acting between the tip and sample as a function of the tip bias can then be calculated using the method of images[20]. Our calculation is calibrated to the tunneling current by examining previously published data,[21] and it is corrected for the motion of the tip relative to the sample surface based on measurements of stationary graphene on Cu foil.[19] The calculated electrostatic force of the STM tip acting on the pristine graphene membrane as a function of the applied bias is then used to transform our height-voltage curves [shown in Fig. 3(c)] into the force-height curves shown in Fig. 3(d). These force curves illustrate in a simple format that 1-3 nN is needed to stretch the sample. By examining the area under each curve, the total energy expended by the electrostatic force is estimated. The area under each curve (shaded area) is calculated to be between 30-60 eV.

EM-STM measurements were then performed on the $SnO_2$ functionalized freestanding graphene, and the results are shown in Fig. 3(e). Five height profiles were taken sequentially in time starting with the bottom trace (i.e., lowest current) and going up. The first three curves (top two curves will be discussed later) show the height increases with current setpoint, and they produced similarly shaped profiles to the pristine freestanding graphene. Astonishingly, however, the height change is now 250-375 nm during these measurements, which is more than a factor of ten greater than pristine graphene. Similar to the freestanding graphene these line profiles can be repeated an unlimited number of times, and each measurement gives a similar result. The force acting on the $SnO_2$ functionalized graphene was calculated the same way as before, and it is shown as a function of the sample height in Fig. 3(f). The electrostatic force is similar to before, but because the displacements have increased by a factor of ten, the energies associated with them have increased dramatically from 28 to 1720 eV for a current of 0.10 nA, for example.

Unlike the lower setpoint currents, the EM-STM measurement taken at 2.00 nA induced an irreversible change in the functionalized graphene and is shown in Fig. 3(e). After that measurement was taken, subsequent EM-STM measurements resembled the 0.10 nA profile offset to the top of Fig. 3(e). This profile is similar to the pristine graphene in that the total height change is now ten times smaller. Also, the line profile was offset to the top of the plot because the overall position of the STM tip moved to this new vertical location. Only if the STM tip was moved to a new sample location more than one micron away, did the large displacement characteristics occur again. The large tunneling current induces local and irreversible changes.

We believe that the large current locally heats the surface, and the nanoparticles are detached, leaving behind a region of pristine graphene underneath the tip.

Our model that explains the EM-STM results for the graphene-nanoparticle system begins with the tendency of graphene to wrap around objects when placed in solution. Two separate groups have been able to exploit this behavior to encapsulate bacteria[22] and Pt nanoparticles[23] respectively. In a similar way, we believe that the freestanding graphene in a solution of $SnO_2$ will naturally wrap around and encapsulate the nanoparticles. Also, due to the negative thermal expansion coefficient of graphene, as the system cools to room temperature the nanoparticles will shrink as the graphene expands to aid the wrapping process. With this model, one can argue that the movement of the graphene due to the STM will gradually and systematically peel away the graphene from the nanoparticles. An illustration of this concept is shown in the lower section of Fig. 3(b). Before the voltage is increased, the graphene is contracted, and the nanoparticles are encapsulated. After the voltage is increased, the graphene is stretched, with the excess graphene coming from the unwrapping of the nanoparticles. Thus, as the tip moves back and forth the graphene folds and unfolds around the nanoparticles. Ultimately, as the STM tip is displaced with higher setpoint currents, the local area would be heated, and the nanoparticles would be easily detached.

In conclusion, using a chemical solution route freestanding graphene was functionalized with $SnO_2$. Chemical analysis using EDS confirmed the presence of the constituent elements, and SEM images revealed that uniform coverage is achieved. The elastic properties of the $SnO_2$ functionalized freestanding graphene were explored using EM-STM. The functionalized freestanding graphene was found to displace ten times farther than the pristine freestanding graphene under a similar force. However, high-current EM-STM did permanently modify the coated graphene on a local scale. A model that includes the encapsulation of the nanoparticles by the graphene was presented.

**Acknowledgements**

L.D. acknowledges financial support by the Taishan Overseas Scholar program, the National Natural Science Foundation of China (51172113), the Shandong Natural Science Foundation (JQ201118), the Research Corporation for Science Advancement, and the National Science Foundation (DMR-0821159). P.X. and P.T. are thankful for the financial support of the Office of Naval Research under Grant No. N00014-10-1-0181 and the National Science Foundation under Grant No. DMR-0855358.

**Figure Captions**

Fig. 1. Schematic illustration of the chemical functionalization process. (a) 100 mg of $SnCl_2$ and 175 μL of HCl (38%) are mixed in 10 mL of deionized water and sonicated for 10 minutes, forming the $SnO_2$ nanoparticles (represented in green). (b) The 2000-mesh Cu grid overlaid with graphene is placed in the solution containing the nanoparticles for 60 minutes. (c) The freestanding graphene is then washed with deionized water and dried in an oven at 70°C for 12 hours.

Fig. 2. (a) Large scale SEM image of pristine freestanding graphene supported by a 2000-mesh Cu grid. (b) Small scale SEM image of pristine graphene. Darker areas represent regions in which graphene is not present. (c) Large scale SEM image of the freestanding graphene functionalized with $SnO_2$ nanoparticles. (d) Small scale HAADF STEM image of the functionalized graphene with bright areas showing the $SnO_2$ nanoparticles. (e) EDS spectrum of $SnO_2$ functionalized freestanding graphene. The Cu and Al signals result from the support grid and the sample holder respectively.

Fig. 3. (a) Feedback-on, height-voltage curves taken with the STM over an Au substrate using three different setpoint currents. (b) Top schematic illustrates the STM tip performing the EM-STM measurement on a pristine freestanding membrane with a low bias (left side) and a high bias (right side) between the tip and the sample. Bottom schematic illustrates EM-STM on functionalized freestanding graphene. Notice how the nanoparticles (green) are unwrapped and therefore allow for more graphene displacement. (c) EM-STM data of pristine freestanding graphene with tip height plotted against bias voltage for three different setpoint currents. (d) Calculated electrostatic force plotted against graphene height for the data shown in (c). (e) Five different EM-STM data sets taken on $SnO_2$ functionalized graphene and presented chronologically starting first with the bottom trace (i.e., lowest current) and moving up. (f) Calculated electrostatic force plotted against sample height for the data shown in (e).

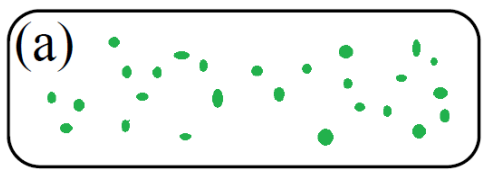

100 mg of SnCl$_2$ in 10 mL H$_2$O
Add 175 μL HCl (38%)
Sonicate 10 mins

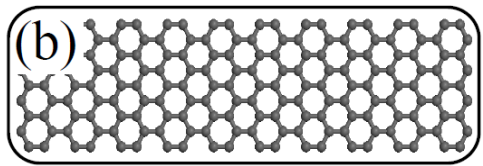

Cu grid with graphene
In solution for 60 min.

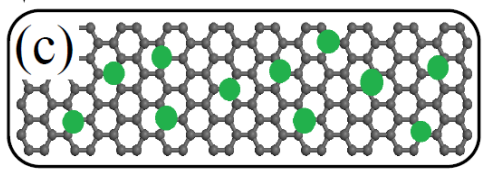

Remove grid from solution
Wash with deionized water
Dry at 70˚C for 12 hours

Fig. 1 by Dong *et al*.

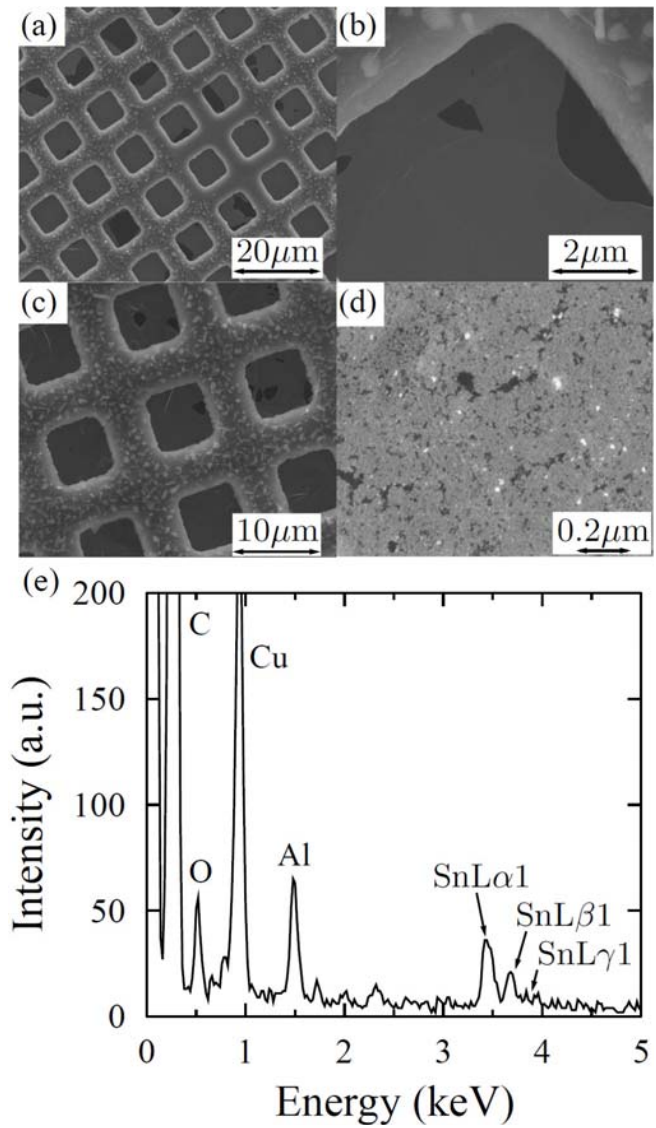

Fig. 2 by Dong *et al*.

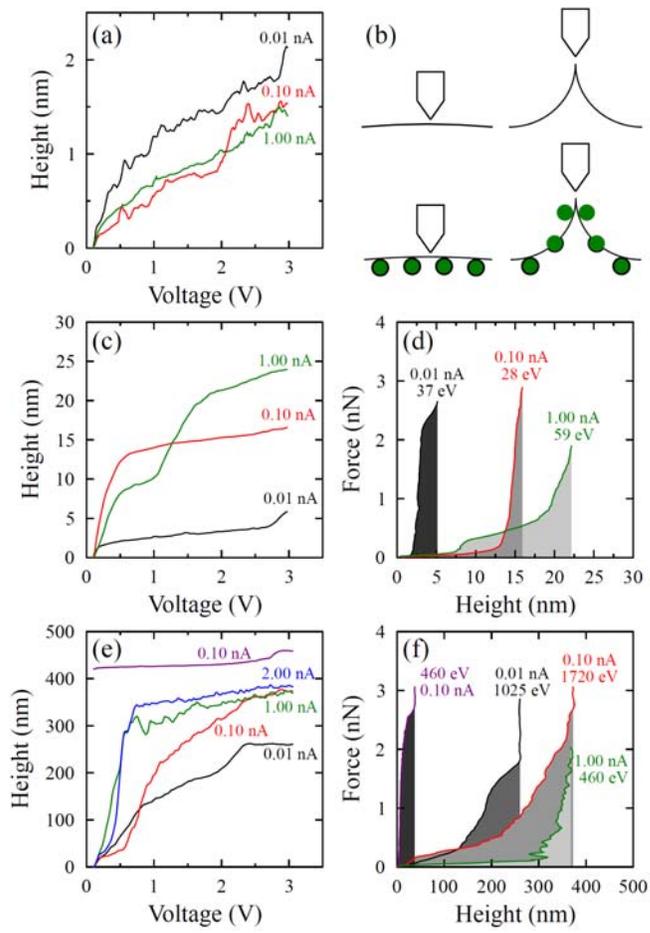

Fig. 3 by Dong *et al*.